# (Un)conscious Bias in the Astronomical Profession: Universal Recommendations to improve Fairness, Inclusiveness, and Representation

Alessandra Aloisi[1]* and Neill Reid[1]

[1]*Space Telescope Science Institute, 3700 San Martin Drive, Baltimore, MD 21218, USA*

**Abstract**

(Un)conscious bias affects every aspect of the astronomical profession, from scientific activities (e.g., invitations to join collaborations, proposal selections, grant allocations, publication review processes, and invitations to attend and speak at conferences) to activities more strictly related to career advancement (e.g., reference letters, fellowships, hiring, promotion, and tenure). For many, (un)conscious bias is still the main hurdle to achieving excellence, as the most diverse talents encounter bigger challenges and difficulties to reach the same milestones than their more privileged colleagues. Over the past few years, the Space Telescope Science Institute (STScI) has constructed tools to raise awareness of (un)conscious bias and has designed guidelines and goals to increase diversity representation and outcome in its scientific activities, including career-related matters and STScI sponsored fellowships, conferences, workshops, and colloquia. STScI has also addressed (un)conscious bias in the peer-review process by anonymizing submission and evaluation of Hubble Space Telescope (and soon to be James Webb Space Telescope) observing proposals. In this white paper we present a plan to standardize these methods with the expectation that these universal recommendations will truly increase diversity, inclusiveness and fairness in Astronomy if applied consistently throughout all the scientific activities of the Astronomical community.

## 1. Why Diversity is important in Science

Diversity matters both in the workplace and in everyday life! Diversity allows for a wider range of people, skills, talents, and perspectives to be involved. When this happens, fewer concepts are taken for granted, more questions are raised in discussions, and more viewpoints are taken into account in devising solutions. The key to a breakthrough is often the ability to see the problem from different perspectives, not simply "being smart" (Page, 2007). When groups of highly talented people work together to solve a hard problem, what matters most in achieving an output that is bigger than the sum of the individual contributions is their diversity, rather than their individual abilities (Ely & Thomas 2001; Parrotta, Pozzoli, & Pytlikova 2011; Ostergaard, Timmermans, & Kristinsson 2011). In other words, diversity is not separate from increasing overall quality, rather it is essential to achieve it.

Diversity is particularly important in science where knowledge progression is made in teams and is based on breakthroughs (Temm 2008). According to Kenneth Gibbs, Jr., Ph.D. (2014), "Diversity in science refers to cultivating talent and promoting the full inclusion of excellence across the whole social spectrum" in a way that includes people of, e.g., all genders, races, and backgrounds just to mention a few key aspects of individuals. This means that if we want excellence in science we need to continually focus on hiring, cultivating, and retaining talent that would not be otherwise accessible (Carrell, Page, & West 2010).

## 2. Science and The Paradox of Meritocracy

Many may view diversity efforts in science as intrinsically contrary to the ideal of meritocracy, a system that rewards a combination of ability and effort in such a way that according to the Merriam-Webster dictionary definition, "the talented are chosen and moved ahead on the basis of their achievement".

However, it is necessary to acknowledge the existence of the so-called "Paradox of Meritocracy". When a company's core values emphasize meritocratic values, those in managerial positions award, for example, larger rewards (salary, promotions), to male employees than to equally performing female employees (Castilla 2008; Castilla & Bernard 2010). Those who believe they are the most objective (such as scientists) may actually exhibit the biggest bias as they do not feel it is necessary to monitor and scrutinize their own behavior.

### 2.1. Simulating Gender Bias Effects

The paradox of meritocracy and its interplay with gender bias has been modelled through computer simulations by Martell, Lane, & Emrich (1996). The cumulative effects of gender bias on career and organizations is presented in Figure 1 in a simulation performed in Spring 2018 from the website http://doesgenderbiasmatter.com/ (unfortunately no longer available). In the simulation, the organization starts with a 1:1 gender ratio at each level (from entry level 1 to executive level 8). All employees go through a performance evaluation cycle twice a year for 10 years, with the performance review scores randomly generated for each employee. In the specific simulation presented in Figure 1, a 5% increase was artificially introduced in the performance review scores for men (on a performance scale from 1 to 100, with a typical value for very successful employees around 70, a 5% increase would yield 73.5 which is well within the errors of a performance assessment). The simulations consider a 15% turnover at all levels, with new positions filled with the highest-ranking performers from the preceding level. In 10 years, the organization ends up unbalanced with less women in leadership positions, i.e., only 30% and 25% respectively at the senior and executive levels (levels 7 and 8).

In summary, organizations need to continuously counteract (un)conscious bias effects in hiring, promotion and leadership selection in order to level the playing field not only for women, but even more for other under-represented groups.

While this simulation reproduces what happens in an organization, this reality is not too far off compared to the situation that women face at every level of their scientific career in academia. This result holds for other under-represented minorities with effects that accumulate if an individual is at the intersection of multiple under-represented dimensions (e.g., women of color).

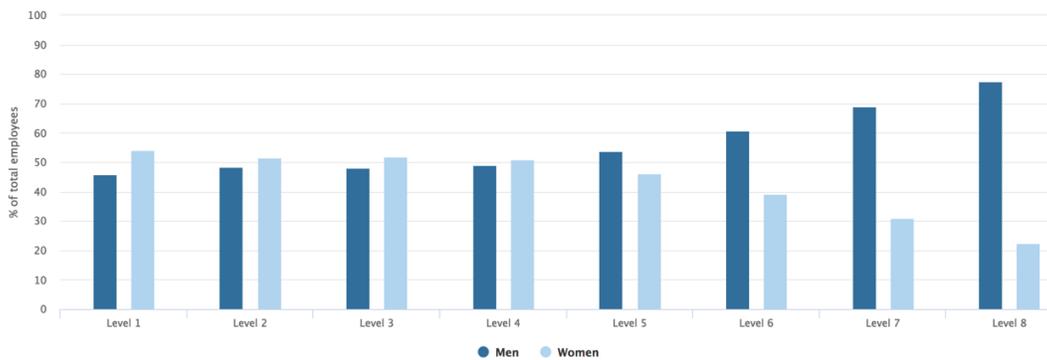

**Figure 1**: Changes in gender ratio when a 5% bias is introduced in favor of men vs. women in the simulation of a generic organization starting with a 1-1 gender ratio at each level and run over a 10-year period.

## 3. Deconstructing Bias

As scientists, we like to believe that we are all objective. After all, we have been trained to handle data objectively, and to apply the rigor of scientific method to everything we do. While it is tempting to think that only a minority of uninformed people are biased, the truth is that we are all biased. We all perceive and treat people based on their social groups (ethnicity, gender, sexual orientation, etc.; see, e.g., Valian 1998 and Fiske 2002).

In our everyday modern life, we are continuously bombarded with a high volume of information and our brain relies on schemas to make sense of the world around us, especially when quick decisions are required. Schemas can be compared to "cognitive shortcuts" used to increase efficiency in navigating situations. They are conceptual frameworks that help our brain to anticipate what to expect from experiences and situations. We are often not aware that we have schemas, but our brain relies on them particularly when time is short and pressure to make a quick decision is high.

These cognitive roadmaps for quickly processing and categorizing information, are built and enforced over a lifetime. They are created through exposure, i.e., direct contact (e.g., personal experiences), or second-hand contact (e.g., through upbringing, books, social media and popular culture). They are widely culturally shared: e.g., both men and women hold them about gender, and both whites and people of color hold them about race/ethnicity.

Biases are schemas, (un)conscious hypotheses and expectations of others based on their group membership that influence our judgement and how we interact with them (regardless of our own group). These schemas include both "attitudes" - those general gut feelings towards a category (e.g., like vs. dislike) - and "stereotypes" - more specific associations between a category and a particular trait (e.g., Asians are good at math; women are natural caregivers). Attitudes and stereotypes introduce bias in the decision-making process, because they represent information that diverges from a neutral point. In particular, "explicit (conscious) biases" are attitudes and stereotypes that we are aware of and constantly check for accuracy, appropriateness, and fairness to self-correct as needed. "Implicit (unconscious) biases" are instead attitudes and stereotypes that we are not aware of and we cannot identify through direct introspection, but that are still influencing our actions and decisions, sometimes at a great extent.

Ultimately, biases are errors in the decision-making, and for this reason it is extremely important to be aware of them and mitigate their effects.

## 4. How to bring Unconscious Bias to light

Founded in 1998 by three scientists (T. Greenwald, M. Banaji, and B. Nosek), "Project Implicit" is a non-profit organization and international collaboration among researchers focused on providing new ways of understanding attitudes, stereotypes, and other hidden biases that are outside of active awareness and control, but still influence our perception, judgment, and action.

The Implicit Association Test (IAT) is the key tool developed by this organization (see https://implicit.harvard.edu/implicit/). It is a web-based series of exercises that uses the reaction time for concepts associations as the basis for the test. The IAT discovered a significant degree of implicit bias among those tested on several factors including race, gender, sexual orientation, and national origin. It is a particularly useful tool as, when taken, it reveals our implicit attitude towards the factors above mentioned. For example, we may believe that both men and women are equally associated with science, but the test may bring to the surface that we associate men with science more than we associate women, and this belief might be very frequently found.

## 5. Unconscious Bias Happens

Schemas do lead to unconscious bias and affect evaluation of individuals belonging to different groups in two distinct ways (Figure 2): (1) they affect the "standard" we use to evaluate the performance of individuals (Biernat & Manis 1994), and (2) they affect the "criteria" we use to judge the performance of individuals (Valian 1999).

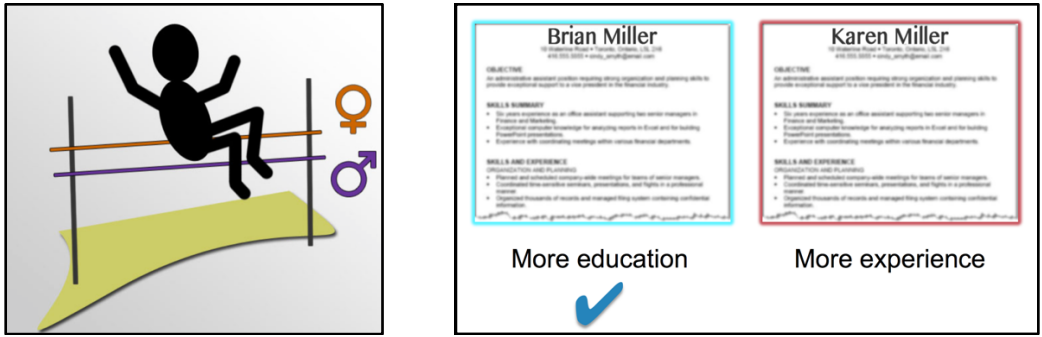

**Figure 2**: The two different ways in which unconscious bias affects evaluation of individuals (from "Fostering Women's Leadership", Davies' keynote presentation at "Building a Culture of Women Leaders" organized by the National District Attorney Association (NDAA) on 20 July 2015). If gender bias is considered, women are held to higher "standards" than men *(left),* or different "criteria" are used from a hiring committee to evaluate a candidate, i.e., experience for a woman versus education for a man *(right)*.

In Astronomy, unconscious bias affects every aspect of the astronomical profession, from the most scientific activities such as invitations to join collaborations, committee memberships, peer-review processes for articles as well as proposals and grants, and invitations to attend and speak at conferences, to activities more strictly related to academic career advancement, such as reference letters, applications for fellowships, hiring, promotion, and tenure.

Some examples of unconscious bias, drawn from science as well as other disciplines, are given below. While not exhaustive, these examples all contain lessons that we should learn and translate into new universal policies, if we want to make the field of Astronomy more inclusive and fairer.

### 5.1. Blind Music Auditions

The most famous example of how unconscious bias affects evaluations is gender bias in orchestra auditions and hires. By analyzing the records of more than 14,000 individuals who auditioned for 8 major US symphony orchestras from mainly 1970 to 1996, Golding & Rouse (2000) found that the adoption of 'blind" auditions with a screen to hide the player's identity from the judging committee (and later on a carpet to conceal the footsteps that could unravel the sex of the candidate), increased the probability that a woman would advance from preliminary rounds to subsequent rounds by 50%. Roster data (lists of orchestra personnel from programs) from 11 major US orchestras show that the switch to blind auditions accounted for ~ 30% of the increase in the proportion of women among new hires between 1970 and 1996.

### 5.2. Gender Bias in Evaluations of Curricula Vitae

Steinpreis, Anders, & Ritzke (1999) studied the effects of gender bias in academia through the evaluation of curricula vitae (CVs) by outside reviewers and search committee members. For this purpose, they used the same CV of a real-life scientist at two different stages of her career (job search and tenure), and assigned a male or a female name to each of the two CVs. What they found is that when evaluating identical application packages for an assistant professorship, male and female psychology professors preferred 2:1 to hire a male candidate (Brian Miller) over a female candidate (Karen Miller). This is consistent with earlier studies that have shown how department heads were significantly more likely to hire female candidates at the assistant professor level and male candidates with identical records at the associate professor level (e.g., Fidell 1970). Similar results were found by Moss-Racusin et al. (2012) for evaluations of graduate student application materials.

Steinpreis, Anders, & Ritzke (1999) also showed that when evaluating identical packages at a more experienced level (tenure), there was no effect of gender on the respondents' recommendations for hiring or granting tenure. However, reservations were expressed four times more often when the identifier was Karen rather than Brian.

### 5.3. Race Bias in Evaluations of Resumes

Bertrand & Mullainathan (2004) studied the effects of race bias in the labor market by sending out fake resumes with randomly assigned white-sounding names (e.g., Emily or Greg) and African-American names (e.g., Lakisha or Jamal) to help-wanted advertisements in Chicago and Boston newspapers. What they found is that Emily or Greg needed to send about 10 resumes to get a callback, compared to Lakisha or Jamal who needed to send about 15 resumes, a significant 50% impact.

On average, Lakisha or Jamal needed 8 more years of experience to get as many callbacks as Emily or Greg. The higher the resume quality, the larger the gap between callbacks for Emily/Greg and Lakisha/Jamal.

### 5.4. Gender Bias in Letters of Recommendation for Academia

A couple of peer-reviewed articles have investigated how gender bias affects letters of recommendations for jobs in academia, specifically postdoctoral positions (Dutt et al. 2016) and faculty positions (Madera, Hebl, & Martin 2009) at US Universities and Institutes. What these studies have found are profound differences in the way male and female applicants are described. The tone of the letters was independent of the writer's gender.

Male applicants are more likely to get excellent letters. They are described in terms of action-oriented characteristics such as "confident", "assertive", "aggressive", "ambitious", and "independent". In addition, terms like "brilliant scientist", "trailblazer", and "one of the best students I've ever had" are often used.

Female applicants, on the other hand, are more likely to get good letters. In these letters, they are more often described in terms of relationship-building characteristics such as "nurturing", "caring", "kind", "agreeable", and "warm". Terms like "solid scientist doing good work", "highly intelligent", and "very knowledgeable" are frequently used in this case.

It is easy to conclude from these studies that women are tendentially not described in terms of the leadership skills necessary to excel in science, and are thus penalized when applying for jobs or coming up for tenure.

### 5.5. Biases in Peer-Review Process

Peer review is at the basis of science. Yet, (un)conscious biases affect such process in a significant way. Several studies have been undertaken to highlight these issues, and single-blind reviews (where the reviewer knows the names and affiliations of the authors), have been replaced by dual-blind reviews (where neither author nor reviewer identity is disclosed) in an attempt to mitigate bias effects.

Dual-blind review was introduced in 2001 by the journal *Behavioral Ecology.* Following this change, a significant 8% increase in first-author papers by female authors, and a corresponding 8% decrease in first-author papers by male authors, was recorded (Budden et al. 2008).

In a more recent study, Tomkins, Zhang & Heavlin (2017) performed a controlled experiment where contributions for a highly prestigious conference in computer science were simultaneously assigned to an equal number of selection committee members for single- or dual-blind reviews. For this discipline, research first or exclusively appears in peer-reviewed conferences rather than journals, so this experiment is indicative of the competitive peer-review process in other STEM fields, including Astronomy. The experiment highlighted a significant preference of single-blind reviewers to favor famous authors and authors from prestigious institutions.

### 6. Tackling (Un)conscious Bias at STScI

The Space Telescope Science Institute, operated by the Association of Universities for Research in Astronomy (AURA) on behalf of NASA, is the science operations center for the Hubble Space Telescope (HST), the James Webb Space Telescope (JWST) and the Wide-Field Infrared Space Telescope (WFIRST). STScI also manages and operates the Mikulski Archive for Space Telescopes (MAST), and is the data management center for both the Kepler mission and the Transiting Exoplanet Survey Satellite (TESS).

STScI houses a full-time scientific staff (the equivalent of faculty in academia) who help ensure that the NASA missions supported by the institute perform at their best. The more than 100 researchers employed by AURA as well as the European Space Agency (ESA) and the Canadian Space Agency (CSA), perform original research in a broad range of Astrophysical topics including solar system, exoplanets, star formation, galaxy formation and evolution, and cosmology.

The institute, similarly to other astronomical departments, has several committees that manage a variety of scientific activities ranging from recruitment, evaluation, and promotion of research staff, to organization of the weekly colloquium and the yearly scientific symposium. In order to mitigate the effects of (un)conscious bias, over the past few years STScI has constructed tools to specifically raise awareness of (un)conscious bias and has designed guidelines and goals to increase diversity representation and outcome in its scientific activities.

### 6.1. Diverse Committees for a Diverse Outcome

One of the most powerful initiatives was born out of the Women in Astronomy Forum (WiAF) at STScI (De Rosa et al. 2019). Based on the idea that a more diverse committee ensures a more diverse outcome (Casadevall & Handelsman 2014), WiAF proposed uniform guidelines on binary gender representation goals for each scientific committee at STScI and provided recommendations on how to achieve such goals in a homogenous way. The goal is 40% representation in each committee, similarly to the fraction of women in early-career Astronomer positions, with a minimum floor of 27% corresponding to the representation of women in the research staff at STScI. We also created a baseline record based on the past 3-5 years that highlighted the bias present in our past activities, as well as metrics and tools to track progress towards our new goals. While the focus of the current work is on binary gender, we do believe that these best practices will significantly mitigate biases against other under-represented groups.

### 6.2. Tailored (Un)conscious Bias Training

To aid with the goal of increasing representation and fairness, STScI staff have also compiled an extensive presentation on (un)conscious bias, with modules tailored to the specific activities of each scientific committee. This presentation is given at the start of each committee's activity. We are planning to convert this presentation into a video. This will make it easier for committee members to take parts of the bias training specific to their scientific activity and take the training at their own pace.

### 6.3. Newly Adopted Best Practices for Scientific Committees

Considerations from the ad-hoc presentation developed by STScI staff and recent practical experiences have converged into the formulation of the following recommendations as an effective tool to combat (un)conscious bias at every step of the scientific activities performed by each STScI committee:

- Increase awareness of how implicit bias can affect evaluation by taking the IAT tests, particularly those related to gender and race; the results are kept confidential but allow each committee member to become aware of their potential biases (https://implicit.harvard.edu/implicit/).
- Avoid use of global (subjective) judgement. Define instead explicit and objective evaluation criteria, and their relative importance, prior to beginning deliberation.
- In case of lack of information, unconscious bias kicks in more easily. To avoid that, all the materials submitted for each case (application, proposal, or tenure package) must be carefully read.
- Committee members must be mindful of biases in recommendation letters when evaluating applications or promotions with external letter writers.

- Sufficient discussion time must be given to each case (application, proposal or tenure package).

- During group discussions, committee members must be mindful of other committee members and respect their opinions. Chair should actively engage all committee members in the discussion.

- During deliberation, objective criteria must be constantly reminded and specific evidence must be offered in support of judgement.

- Committee members should refrain from identifying individuals by name and/or gender in discussions, but rather use neutral terms like, e.g., "the applicant", "the candidate", or "they" for individuals applying for jobs, and "the proposers" or "the proposal team" when discussing proposals.

- For scientific proposals, committee members should resist the temptation to discuss PI's individual qualifications; focus instead on the scientific and/or technical merits of the project. Proposals do not generally include CVs of the proposing team members. Any discussion of the proposing team's qualifications risk being influenced by biases (discussion based on perceived qualifications, not quantitative metrics) and extent of the proposer's network (networks of underrepresented individuals are smaller and have fewer connections to high-status scientists).

In addition, for hiring activities including selection of postdoctoral fellows and recruitment of research staff, the following recommendations should also be considered:

- The number of women and minorities in the applicant pool should be increased through active and open recruitment.

- The applicant pool demographics should be reviewed with Human Resources at every step of the selection process (triage, phone interviews, in-person interviews, offers).

- Structured interviews should be used - same questions and same format for all candidates.

Even if a scientific committee has achieved the expected representation, has been trained about (un)conscious bias, and has followed all the steps outlined above, the outcome may still end up being biased. Ex-officio members should be present at committee deliberations to ensure that the best practices are followed. If the end result is still affected by bias, the recommendation is to raise awareness, discuss the biased outcome and redo the process.

### 6.4. Dual-Anonymous Review Process for HST Proposal Selection

STScI manages the activities required to implement the highly competitive selection of the science program of HST and in the near future JWST. In the past, the HST proposal selection has shown evidence of gender bias. By analyzing data collected between 2001 and 2012, Reid (2014) found that the success rates for proposals with women as principal investigators (PIs) were lower (19%) than the success rates of their male counterparts (23%).

In order to counteract this gender bias as well as other biases related to factors such as race, career stage, institutional size, and geographic origin, STScI introduced in 2018 a dual-anonymous review process where names of both proposers and reviewers remain unknown to all until after the end of the process. This first-of-its-kind process for proposals selection, resulted in women PIs having a higher success rate compared to men (8.7% versus 8.0%, respectively) for the very first time in 18 years (Figure 3). This result was so encouraging that NASA is now implementing dual-anonymous reviews for all its facilities (Witze 2019).

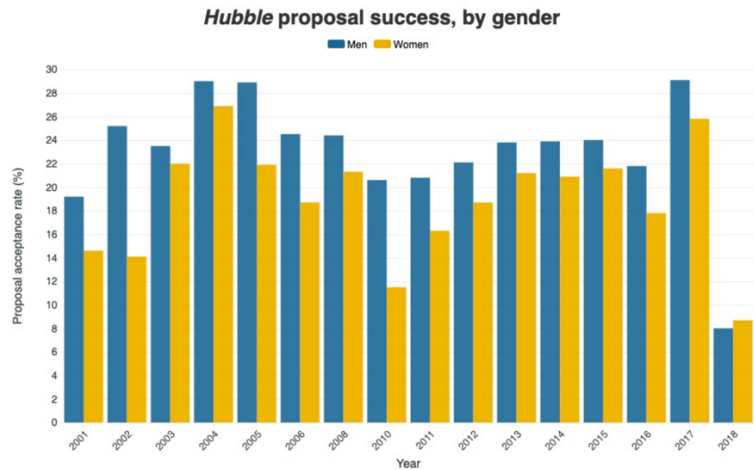

**Figure 3**: HST proposal acceptance rate by gender. From year 2001 to year 2017, there is a systematically higher success rate for men PIs than female PIs. This gender success rate drastically changed in 2018, when dual-anonymous peer-review process was introduced for the very first time in proposal selections to combat (un)conscious bias (adapted from Strolger & Natarajan 2019).

## 7. Recommendations and Conclusions

STScI has invested time and effort in the last ten years to mitigate the effect of implicit bias across its activities, to level the playing field for ALL its constituencies. This includes scientific activities (e.g., committees to generate speaker lists, colloquium series, Symposia), career advancement related activities (e.g., recruitment and personnel committees, hiring, promotion, and tenure), and peer-review activities (time allocation, grant allocation, fellowships).

Based on documented evidence in the literature and past and on-going STScI efforts to implement best practices to counteract (un)conscious bias, and encouraged by the positive results that these measures are achieving, we recommend the universal adoption of the following practices that will benefit all underrepresented groups in Astronomy.

- Introduce **bias training** as a necessary preamble to all activities that involve a committee where discussion is necessary and selective decisions are expected. The STScI experience is that bias training greatly improves bias awareness, and makes the subsequent discussion fairer and less prone to gross bias effects.

- Introduce **explicit and objective criteria** before starting the evaluation process and consistently use them throughout. Offer specific evidence in support of judgement. Advertisements for jobs should explicitly state the criteria used to select candidates.

- **Track progress**. A significant step in the process is to establish an institutional baseline of representation on committees in relation to their outcomes. This will allow each institution to verify whether the introduced measures are working and will document progress toward reaching the overall goal of a leveled and fair playing field.

- Introduce **dual-anonymous peer-review process**, not only for allocating **telescope time**, but also when disbursing **grants** and **funds**. Preliminary findings from the HST proposal selection processes in 2018 and 2019 show encouraging signs in several areas. As an added bonus, reviewers remarked how the dual-anonymous process allowed them to focus only on the science and not on the scientist, in a process that some of them defined "liberating".

- Introduce **dual-anonymous processes in recruitment**. While this idea is still in its infancy, we recommend that institutions consider implementing pilot projects in a concerted fashion (maybe coordinated by the AAS). Our expectations are that this would significantly improve the way short lists are made and candidates are recruited and would greatly benefit minority constituencies. This would also allow astronomers to review and revise (and maybe delete altogether) the process of submitting reference letters that are often biased. We advocate exploring possibilities at institutional level and disseminate results nationwide.

Our expectation is that these universal recommendations will truly increase diversity, inclusion and fairness in Astronomy if applied consistently across all the scientific activities of the Astronomical community and all institutions. We also advocate sharing and communicating experiences and results from the implementation of these practices widely so that the entire astronomical community can benefit from learning of what has worked at different institutions. This effort can only succeed if the approach is shared in a collective fashion by the entire Astronomical community.

**Acknowledgements**

The authors thank STScI leadership, in particular the director Ken Sembach and the deputy director Nancy Levenson, for fully supporting initiatives related to diversity and inclusiveness.  We acknowledge Antonella Nota for her suggestion to write this paper based on recent new policy implementations at STScI, and for having pioneered some of the initiatives described above.  The authors also thank Sheryl Bruff, Sylvia Baggett, Karrie Gilbert and the Diversity, Culture, and Respect Working Group of the Instruments Division at STScI for inspiring several parts of this paper. Similarly, Joan Schmelz and the ADVANCE grant program at The University of Michigan "STRIDE" led by Dr. Abigail J. Stewart are also acknowledged.